\RequirePackage{fix-cm}
\documentclass{svjour3}   % onecolumn (standard format)
\smartqed  % flush right qed marks, e.g. at end of proof
\usepackage{cite}
\usepackage{amsmath,amssymb,amsfonts}
\usepackage{algorithmic}
\usepackage{graphicx}
\usepackage{textcomp}
\def\BibTeX{{\rm B\kern-.05em{\sc i\kern-.025em b}\kern-.08em
    T\kern-.1667em\lower.7ex\hbox{E}\kern-.125emX}}
    
\usepackage{epstopdf} %converting to PDF
\usepackage{color}
\usepackage{subfig}
\usepackage{times}
\usepackage{mathptm}
\usepackage{booktabs}
\usepackage{float}
\usepackage{lipsum}
\usepackage{mathtools}
\usepackage{cuted}
\usepackage{cite}

\begin{document}

\title{PAPR Reduction of FBMC Signals Based on Uniform and Linear PDF Companding Schemes}
\author{Srinivas Ramavath \and Umesh Chandra Samal}


\institute{Srinivas Ramavath \at 
              \email{srinivas2012iitg@gmail.com}   \\
                      School of Electronics Engineering, Kalinga Institute of Industrial Technology (KIlT) University
           \and
           Umesh Chandra Samal \at
           \email{umesh.samalfet@kiit.ac.in}\\
              School of Electronics Engineering, Kalinga Institute of Industrial Technology (KIlT) University
}

\date{Received: date / Accepted: date}

\maketitle

\begin{abstract}
In this paper, two new companders are designed to reduce the ratio of peak to average power (PAPR) experienced by filter bank multicarrier (FBMC) signals. Specifically, the compander basic model is generalized, which alter the distributed FBMC signal amplitude peak. The proposed companders design approach provides better performance in terms of reducing the PAPR, Bit Error Rate (BER) and phase error degradation over the previously existing compander schemes. Many PAPR reduction approaches, such as the $\mu$-law companding technique, are also available. It results in the formation of spectrum side lobes, although the proposed techniques result in a spectrum with fewer side lobes. The theoretical analysis of linear compander and expander transform for a few specific parameters are derived and analyzed. The suggested linear companding technique is analytically analysed using simulations to show that it efficiently decreases the high peaks in the FBMC system. \\
Keywords- BER, CCDF, Compander, FBMC, OQAM, PAPR, PSD.
\end{abstract}

% % % %
% % % %
\section{Introduction}
\label{section:Introduction}
The fifth generation (5G) network is to accommodate the envisioned Internet of Things (IoT) services, which requires high data rate ensuring reliability. One of the most challenging problems of upcoming wireless communication technology is a massive interconnection of users and devices for more flexibility and efficiency essential to support heterogeneous system requirements and to use spectrum efficiently as much as possible. To fulfill these requirements, new waveforms based on FBMC techniques have been proposed \cite{samal20195g}. For future wireless communication systems, the FBMC with offset quadrature amplitude modulation is currently gaining a lot of attention \cite{jiang2017oqam, nissel2017filter}. Using a prototype filter, FBMC-OQAM delivers great spectrum efficiency, enhanced time-frequency localisation, and very minimal out-of-band radiation \cite{bellanger2010fbmc}. However, in compared to OFDM signals, FBMC signals have a fundamentally different signal structure: adjacent signal data blocks overlap in FBMC systems, whereas symbols in OFDM systems do not \cite{prasad2004ofdm,cho2010mimo}. Therefore, the PAPR of FBMC systems is quite different from OFDM systems and needs to be reduced for the better performance \cite{ramavath2012analytical}.\\

Signal distortion happens when big peaks enter a nonlinear section of a high-power amplifier (HPA) without being preprocessed \cite{an2018design}. As a result, reducing the signal peak power is a good way to alleviate the high PAPR problem. So for In the literature, many PAPR reduction techniques have been devised \cite{he2017filter,na2017low}. There are three primary categories in which these schemes may be classified:
Techniques for signal distortion, signal scrambling, and coding. Before passing through the HPA, signal distortion methods lower the PAPR of FBMC signals. Filtering and clipping \cite{cho2010mimo}, peak windowing, peak cancellation, and other companding techniques \cite{zhang2017filtered,adebisi2018enhanced,jiang2007nonlinear} are among the most well-known signal distortion techniques.
These approaches considerably reduce the power ratio, but they also create in-band and out-of-band distortion \cite{zakaria2016theoretical}, resulting in BER degradation. The alternative option is to rearrange the data symbols without altering the way signals are generated. such as the selective mapping (SLM)\cite{bulusu2014reduction,ji2014semi,li2014partition}, tone reservation (TR)\cite{lu2012sliding}, partial transmission sequence (PTS) \cite{lee2016low,deng2017modified,ye2013papr} and active constellation extension (ACE) \cite{van2014papr,yang2005ace,liu2019low,cui2019peak}. It should be emphasised that the two categories listed above are rarely antagonistic. Modifying and extending the PAPR reduction methods for FBMC signals to Companded FBMC signals is a natural way \cite{ramavath2021theoretical}.\\

The FBMC system is explored in this study\cite{chunkath2021constrained,geetha2020performance,srivastava2020hybrid}, which includes two unique companding operations at the transmitter. The authentic signal amplitude is changed into a predetermined distribution in this work to reallocate the statistics of the compounded signal more freely. The uniform and nonlinear companding techniques are presented in particular to deal with signal distortion. The theoretical and simulation result show that the proposed techniques suppresses the PAPR more effectively. The nonlinear companding technique achieves a better bit error rate (BER) than other companding techniques.\\

The rest of this paper is laid out as follows. The FBMC system model and problem description are described in Section II. In section III, we derive a theoretical analysis of the proposed schemes. In section IV, the PAPR and BER performance are simulated and compared with the previoulsy existing $\mu$-law companding scheme. Finally, it is conclude in Section V.

 %=====================================================================================================================
\section{FBMC SYSTEM Model}
\begin{figure*}[ht]
\includegraphics[scale=0.41]{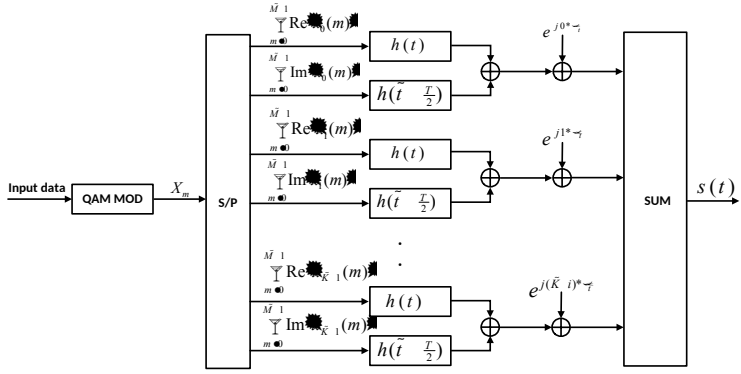}
\caption{The FBMC-OQAM system transmitter}.
\end{figure*}
\subsection{Transmitted Signal}
In FBMC system, the input $m^{th}$ complex data symbol for $k^{th}$ subcarrier can represented as
\begin{equation}
x_{k}(m)=a_{k}(m)+jb_{k}(m) ,\,\,\,\,\,\,0 \le k \le K - 1,\,\,\,\,\,\,0 \le m \le M - 1
\end{equation}
The real and imaginary components of the $m^{th}$ symbol for the $k^{th}$ subcarrier are $a_{k}(m)$ and $b_{k}(m)$, respectively. 
For every K sub-carriers, the $m^{th}$ symbols input data block $x(m)$ is provided by 
\begin{equation}
x(m)=[x_{0}(m),x_{1}(m),.....,x_{K-1}(m)]^{T}
\end{equation}
In the time domain, the real and imaginary components of each symbol are tagged by $\frac{T}{2}$, which is half of the data block period ($T$). 
The data blocks are modulated using $K$ distinct sub-carrier modulators, whose carrier frequencies are spaced by the inverse of the symbol period $(\frac{1}{T})$. The $M$ data block modulated signal generated by the OQAM-FBMC is represented as 
\begin{equation}
S(t)=\sum_{k=0}^{K-1}\sum_{m=0}^{M-1}S_{m}^{k}(t)
\end{equation}
The FBMC system $k^{th}$ sub-carrier and $m^{th}$ symbol signal is expressed as
\begin{equation}
S_{m}^{k}(t)=\left [ Re\left ( x_k \right )h(t-mT)+Im\left ( x_k \right )h(t-mT-\frac{T}{2}) \right ]e^{jk\varphi_{t} }\
\end{equation}
where $\varphi_{t}=\frac{2\pi}{T}t+\frac{\pi}{2},\,\,\,\,\,\,0 \le t \le \left ( M+\alpha-\frac{1}{2} \right )T$ and the time domain impulse response with finite length $\alpha T$ is denoted by $h(t)$. 
Because the prototype filter response in FBMC-OQAM systems has a longer time period than $T$, surrounding data signal blocks overlap. 
\subsection{Problem Description}
Each symbol in an OFDM system has a length of $N$ and a rate of $\frac{1}{N}$. A set of $N$ orthogonal sub-carriers modulates the OFDM input data complex symbol $X(k)$ for $0\le k\le N-1$, For $0\le n\le N-1$, the output $x(n)$ is represented as  
\begin{equation}
x(n)=\frac{1}{N}\sum_{k=0}^{N-1}X(k)e^{j\frac{2\pi}{N}nk},\,\,\,\,\,\,0 \le n \le N - 1
\end{equation}
There are no overlaps between neighbouring OFDM symbols, and the power ratio of the OFDM system is calculated per symbol \cite{prasad2004ofdm}.
\begin{equation}
PAPR_{OFDM}=\frac{1}{P_{avg}}\,\,\,\,\begin{matrix}
    \begin{matrix}
    max_{n}\\
    0 \le n \le N-1\end{matrix}& 
     \left |x(n) \right |^{2}\end{matrix}
\end{equation}
The FBMC-OQAM signals in the time domain are overlapping with signals from the OFDM system that have been changed. The FBMC modulated signal $s(t)$ can be broken into symbols of length $T$ over which the PAPR can be calculated for precise peak to average power ratio comparisons to 4G systems. The procedure of measuring the PAPR over two symbols of length $T$, each shown by $PAPR_{1}$ and $PAPR_{2}$, is depicted in Fig. 2. We can define the PAPR over length $T$ in a frame based on that concept \cite{cui2019peak}.

\begin{figure}[th]
\includegraphics[scale=0.91]{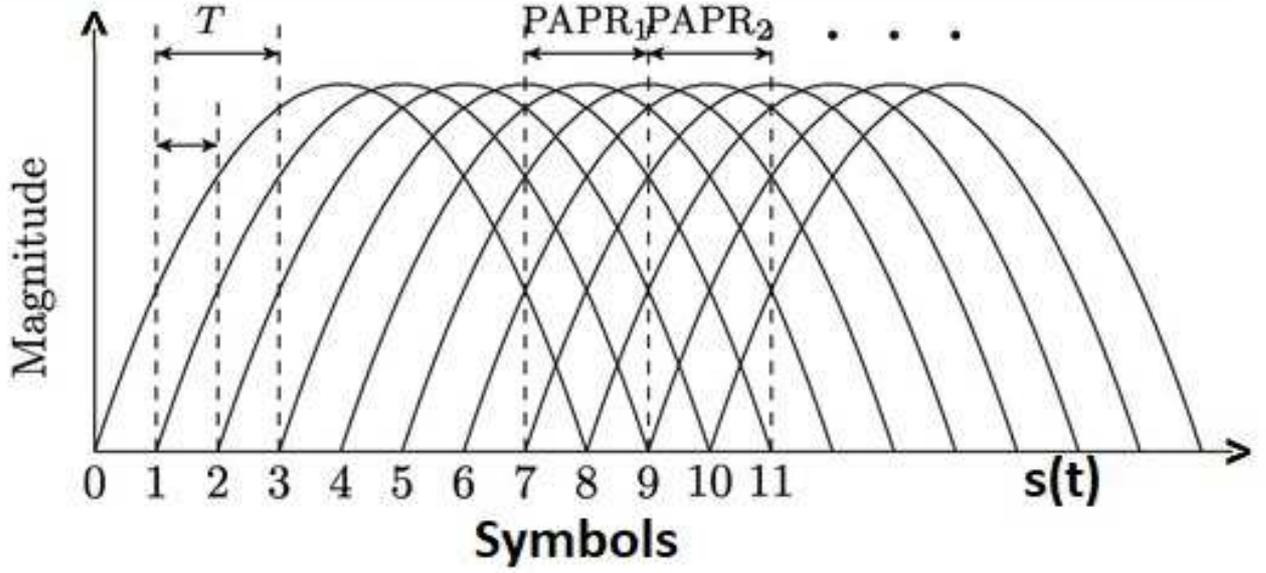}
\caption{FBMC Overlaping symbols}.
\vspace{-0.2cm}
\end{figure}
Each part of $S(t)$ has length $T$ because it is initially split into $M+\alpha $ parts. The power ratio for $S(t)$ in the $i^{th}$ interval is then calculated as  \cite{samal20195g}
\begin{equation}
PAPR_{FBMC}=\frac{1}{P_{avg}}\,\,\,\,\begin{matrix}
    \begin{matrix}
    max_{n}\\
    iT \le t \le \left ( i+1 \right )T\end{matrix}& 
     \left |S(t) \right |^{2}\end{matrix} 
\end{equation}
where $i=0,1,2,.....,M+\alpha -1$.
To simplify Eq.(4), denote $X_{k}(m)=Re\left ( x_k \right )$ if m is even, and $X_{k}(m)=Im\left ( x_k \right )$ if m is odd.
Then, Eq. (4) can be rewritten as
\begin{equation}
S(t)=\sum_{k=0}^{K-1}\sum_{m=0}^{M-1}X_{k}(m)h(t-\frac{mT}{2}) e^{j\left ( \frac{2\pi}{T}t+\phi _{m}^{k} \right )}
\end{equation}
where $\phi _{m}^{k} =\frac{\pi}{2}(m+k)-\pi mk$
hence, the $k^{th}$ subcarrier signals can be expressed as 
\begin{equation}
S^{k}(t)=\sum_{m=0}^{M-1}X_{k}(m)h(t-\frac{mT}{2}) e^{j\left ( \frac{2\pi}{T}kt+\phi _{m}^{k} \right )}
\end{equation}
The FBMC system transmitted signal is given as
\begin{equation}
S(t)=\sum_{k=0}^{K-1}S^{k}(t)
\end{equation}
The variance is $\sigma _x^2$, and $X_{k}(m)$ is a statistically identical independent distribution (i.i.d.). 
The mean and variance of $S^{k}(t)$ is $0$ and $\sigma _{k}^{2}=\sigma _{x}^{2}\sum_{m=0}^{M-1}h\left ( t-{\frac{mT}{2}} \right )^{2}$ respectively.
$S^{k}(t)$ has a mean of $0$ and a variance of $\sigma _{k}^{2}=\sigma _{x}^{2}\sum_{m=0}^{M-1}h\left ( t-{\frac{mT}{2}} \right )^{2}$, respectively. $E{\left [ S^{k}(t) \right ]}$ is uncorrelated with $\sigma _{k}^{2}$ and $k$.  
The $s(t)$ real and imaginary components approach Gaussian distribution with zero mean and variance $\sigma _{s}^{2}=k{\frac{\sigma _{k}^{2}}{2}}$ when $k$ is considered as large as possible by the central limit theorem. 

The $s(t)$ absolute square ($\left |S(t)  \right |^{2}$) follows a central chi-squared distribution, probability density function (pdf) of $\left |S(t)  \right |^{2}$ define as
\begin{equation}
P_{Y}(y)=\frac{1}{2\sigma _{s}^{2}}e^{\frac{-y}{2\sigma _{s}^{2}}}
\end{equation}
Then, suppossing
\begin{equation}
Z=\frac{\left |S(t)  \right |^{2}}{E\left [ \left |S(t)  \right |^{2} \right ]}
\end{equation}
Finally probability density function of Z written as
\begin{equation}
\begin{split}
P_{Z}(z)&=2\sigma _{x}^{2}P_{Y}(2\sigma _{x}^{2}z)\\
&=\alpha _{t}e^{-\alpha _{t}z}
\end{split}
\end{equation}
where $\alpha _{t}=\frac{2}{k\sum_{m=0}^{M-1}h\left ( t-{\frac{mT}{2}} \right )^{2}}$.
The cumulative distribution function (CDF) for Z can be written as follows: 
\begin{equation}
\begin{split}
P(z \le \gamma) &=\int_{0}^{\gamma }P_{Z}(z) dz \\
&=\int_{0}^{\gamma }\alpha _{t}e^{-\alpha _{t}z} dz \\
&=1-e^{-\alpha _{t}\gamma}
\end{split}
\end{equation}
As a result, the PAPR distribution function for FBMC can be expressed as 
\begin{equation}
\begin{split}
P(PAPR \le \gamma) &=P\left ( \bigcap_{i=nT}^{(n+1)T-1}\left |S_{0}(t)  \right |^{2} \le \gamma \right ) \\
&=\prod_{i=nT}^{(n+1)T-1}P\left ( \left |S_{0}(t)  \right |^{2} \le \gamma \right )\\
&=\prod_{i=nT}^{(n+1)T-1} \left ( 1-{e^{-\alpha _{t}\gamma }} \right )
\end{split}
\end{equation}
How PAPR is critical for FBMC system and its requirement, which may cause a severe nonlinear distortion by the power amplifier (PA). Therefore, it is necessary to design an effective technique to mitigate the distortion. With an appropriate choice of the companding function can be reduce more peak power. The proposed method is expected to achieve acceptable PAPR. 
\section{Proposed Schemes}
A unique peice-wise linear function scheme that can effectively improve the PAPR reduction and the BER performance of FBMC transmitted signals is proposed. The nonlinear companding \cite{kondamuri2020non} transform is given by $y_{n}=h(x_{n})$, here $x_{n}$ is the FBMC original signal and the nonlinear companded output signal is $y_{n}$, the peice-wise linear function $h(.)$ only changes the input signals amplitudes.

\subsection{Companding Based on nonlinear function}
Assume the transform $y_{1}=h_{1}(x_{n})$ and the pdf of $y_{1}$, the inflection point and the cutoff point are $c \sigma$ and $A_{c}$ respectively depicts in fig. 3. 
\vspace{-0.5cm} 
\begin{figure}[H]
\center
\includegraphics[scale=0.8]{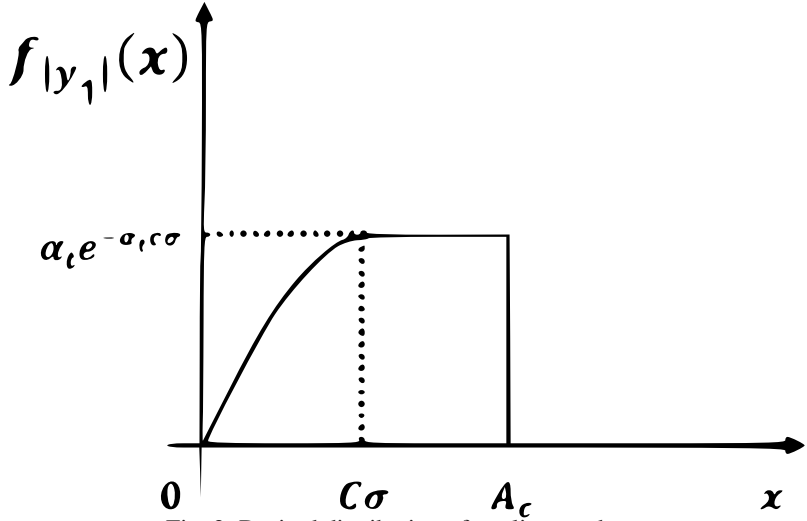}
\vspace{-0.4cm}
\caption{Desired distribution of nonlinear scheme}.
\vspace{-0.5cm}
\end{figure}
The pdf of $\left |y_{1}  \right |$ is same as pdf of $\left |x_{n}  \right |$ between $\begin{bmatrix}
0 & c \sigma
\end{bmatrix}$; In the interval $\begin{bmatrix}
c \sigma & A_{C}
\end{bmatrix}$, the pdf of $\left |y_{1}  \right |$ has a uniformly distributed. The inflection and cutoff point coordinates are ($c \sigma$,$f_{\left | x_n \right |}(x)$ ) and ($A_{c}$,$f_{\left | x_n \right |}(x)$ ), respectively. Therefore, the $\left |y_{1}  \right |$ pdf can be represented as
\begin{equation}
f_{\left | y_1 \right |}(x)=\left\{\begin{matrix}
\frac{2x}{\sigma^{2}}e^{-\frac{x^{2}}{\sigma^{2}}}& 0\leq x\leq c \sigma\\ 
f_{\left | x_n \right |}(c \sigma)&  c \sigma\leq  x\leq A_{c}
\end{matrix}\right.
\end{equation}
The area of the probability density function (pdf) $\int_{0}^{\infty } f_{\left | y_1 \right |}(x) dx =1$ , we can obtain $A=(c+\frac{1}{2c})\sigma$.
The pdf of $|y_{1}(t)|^{2}$ define as
\begin{equation}
f_{|y_{1}(t)|^{2}}(x)=\left\{\begin{matrix}
\frac{1}{2\sigma^{2}}e^{-\frac{y}{2\sigma^{2}}}& 0\leq x\leq c \sigma\\ 
\frac{2c}{\sigma}e^{-c^{2}}&  c \sigma\leq  x\leq A_{c}
\end{matrix}\right.
\end{equation}
Then, suppossing
\begin{equation}
Z_{1}=\frac{\left |y_{1}(t)  \right |^{2}}{E\left [ \left |y_{1}(t)  \right |^{2} \right ]}
\end{equation}
Finally probability density function of $Z_{1}$ written as
\begin{equation}
f_{z_{1}}(z)=\left\{\begin{matrix}
\alpha _{t}e^{-\alpha _{t}z}& 0\leq x\leq c \sigma\\ 
\beta _{t}&  c \sigma\leq  x\leq A_{c}
\end{matrix}\right.
\end{equation}
Where $\beta _{t}=4\alpha _{t}c\sigma e^{-c^{2}}$. The cumulative distribution function (CDF) for $Z_1$ can be expressed as 
\begin{equation}
F_{z_{1}}(z)=\left\{\begin{matrix}
1-e^{-\alpha _{t}z}& 0\leq x\leq c \sigma\\ 
\beta _{t}z&  c \sigma\leq  x\leq A_{c}
\end{matrix}\right.
\end{equation}
Hence, the PAPR distribution function for FBMC can be expressed as 
\begin{equation}
\begin{split}
P(PAPR \le \gamma) &=P\left ( \bigcap_{i=nT}^{(n+1)T-1}\left |S_{0}(t)  \right |^{2} \le \gamma \right ) \\
&=\prod_{i=nT}^{(n+1)T-1}P\left ( \left |S_{0}(t)  \right |^{2} \le \gamma \right )\\
&=\prod_{i=nT}^{(n+1)T-1}\left\{\begin{matrix}
1-e^{-\alpha _{t}z}& 0\leq x\leq c \sigma\\ 
\beta _{t}z&  c \sigma\leq  x\leq A_{c}
\end{matrix}\right.
\end{split}
\end{equation}
Considering the input signal phase, the monotonic increasing function 
$h_{1}$(x) is represented as
\begin{equation}
h_{1}(x)= sgn(x)F_{\left | y_1 \right |}^{-1}(F_{\left | x_n \right |}(x))
\end{equation}
here sgn(x) indicates the signum function. Thus, the nonlinear companding function at the transmitter can be obtained as
\begin{equation}
h_{1}(x)= \begin{cases}
x & \left | x \right |\leq c \sigma\\ 
sgn(x)\frac{\sigma }{c}\left ( \frac{2}{3}-\frac{1}{2}e^{\frac{1}{c^{2}}-\frac{x^{2}}{\sigma ^{2}}} \right ) & \left | x \right |>  c \sigma 
\end{cases}
\end{equation}
At the receiver is the reciprocal of companding transform is
\begin{equation}
h_{1}^{-1}(x)= \begin{cases}
x & \left | x \right |\leq c \sigma\\ 
sgn(x)\frac{\sigma }{c^{2}}\sqrt{\frac{1}{c^{2}}-\frac{1}{c^{4}}ln\left ( \frac{4c\sigma -x}{3c\sigma } \right )}& \left | x \right |>  c \sigma 
\end{cases}
\end{equation}
\subsection{Companding Based on linear function}
Assume the transform $y_{2}=h_{2}(x_{n})$ and the $y_{2}$ pdf, the inflection point and the cutoff point are $c \sigma$ and $A_{c}$ respectively depicts in fig. 4.
 \vspace{-0.3cm} 
\begin{figure}[H]
\center
\includegraphics[scale=0.7]{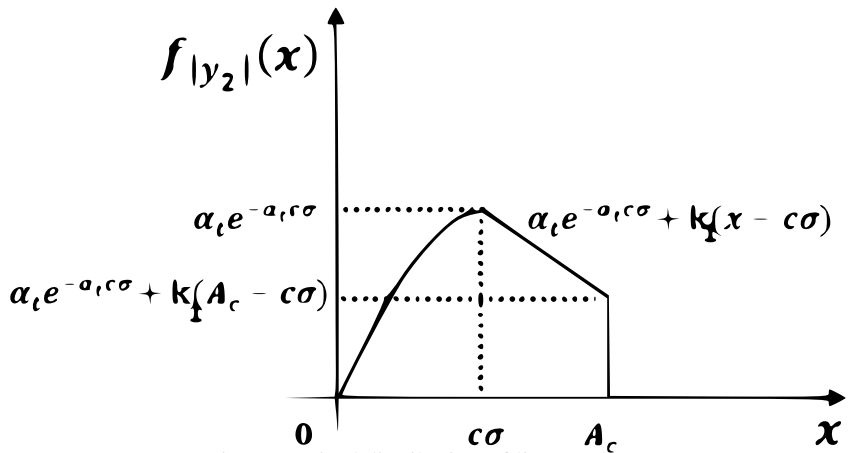}
\vspace{-0.4cm}
\caption{Desired distribution of linear scheme}.
\vspace{-0.5cm}
\end{figure}
The pdf of $\left |y_{2}  \right |$ is identical with $\left |x_{n}  \right |$ between $\begin{bmatrix}
0 & c \sigma
\end{bmatrix}$; In the interval $\begin{bmatrix}
c \sigma & A_{C}
\end{bmatrix}$, the pdf of $\left |y_{2}  \right |$ has linear function with a slope $k_{1}$ distributed. The inflection and cutoff point coordinates are ($c \sigma$,$f_{\left | x_n \right |}(x)$ ) and ($A_{c}$,$f_{\left | x_n \right |}(x)$ ), respectively. Therefore, the $\left |y_{2}  \right |$ pdf can be expressed as
\begin{equation}
f_{\left | y_2 \right |}(x)=\begin{cases}
\frac{2x}{\sigma^{2}}e^{-\frac{x^{2}}{\sigma^{2}}} & 0\leq x\leq c \sigma\\ 
k_{1} x+\alpha _{t}e^{-\delta_{t}}-k_{1} c \sigma & c \sigma\leq  x\leq A_{c}
\end{cases}
\end{equation}
From the definition of CDF $F_{\left | y_2 \right |}(A_{c})=1$, we have obtain
$k_{1}=\frac{4\sigma c^{2}-2A_{c}C+\sigma}{\sigma A_{c}-c \sigma^{2}}e^{-c^{2}}$. 
The probability density function (pdf) of $|y_{2}(t)|^{2}$ define as
\begin{equation}
f_{|y_{2}(t)|^{2}}(x)=\left\{\begin{matrix}
\frac{1}{2\sigma^{2}}e^{-\frac{y}{2\sigma^{2}}}& 0\leq x\leq c \sigma\\ 
\frac{c e^{-c^{2}}}{\sigma \sqrt{y}}+\frac{k_{1}}{2}\left ( 1-\frac{c\sigma }{\sqrt{y}} \right )&  c \sigma\leq  x\leq A_{c}
\end{matrix}\right.
\end{equation}
Then, suppossing
\begin{equation}
Z_{2}=\frac{\left |y_{2}(t)  \right |^{2}}{E\left [ \left |y_{2}(t)  \right |^{2} \right ]}
\end{equation}
Finally probability density function of $Z_{2}$ written as
\begin{equation}
f_{Z_{2}}(z)=\left\{\begin{matrix}
\alpha _{t}e^{-\alpha _{t}z}& 0\leq x\leq c \sigma\\ \\
\frac{c e^{-c^{2}}}{\sigma }\sqrt{\frac{2\sigma _{x}}{z}}+k_{1}\sigma _{x}\left ( 1-\frac{c\sigma }{\sqrt{2\sigma _{x}z}} \right )&  c \sigma\leq  x\leq A_{c}
\end{matrix}\right.
\end{equation}
The $Z_{2}$ cumulative distribution function (CDF) can be written as
\begin{equation}
F_{Z_{2}}(z)=\left\{\begin{matrix}
1-e^{-\alpha _{t}z}& 0\leq x\leq c \sigma\\ 
\frac{2c e^{-c^{2}}}{\sigma }\sqrt{2\sigma _{x}z}+k_{1}\sigma _{x}\left ( z- 2c\sigma \sqrt{\frac{z}{\sigma _{x}}}\right )&  c \sigma\leq  x\leq A_{c}
\end{matrix}\right.
\end{equation}
Hence, for FBMC, the PAPR distribution function can be expressed as 
\begin{equation}
\begin{split}
P(PAPR \le \gamma) &=P\left ( \bigcap_{i=nT}^{(n+1)T-1}\left |S_{0}(t)  \right |^{2} \le \gamma \right ) \\
&=\prod_{i=nT}^{(n+1)T-1}P\left ( \left |S_{0}(t)  \right |^{2} \le \gamma \right )\\
&=\prod_{i=nT}^{(n+1)T-1}\left\{\begin{matrix}
1-e^{-\alpha _{t}z}& 0\leq x\leq c \sigma\\ 
\frac{2c e^{-c^{2}}}{\sigma }\sqrt{2\sigma _{x}z}+k_{1}\sigma _{x}\left ( z- 2c\sigma \sqrt{\frac{z}{\sigma _{x}}}\right )&  c \sigma\leq  x\leq A_{c}
\end{matrix}\right.
\end{split}
\end{equation}
Considering the input signal phase, the monotonic increasing function 
$h_{2}$(x) is represented as
\begin{equation}
h_{2}(x)= sgn(x)F_{\left | y_1 \right |}^{-1}(F_{\left | x_n \right |}(x))
\end{equation}
here sgn(x) indicates the signum function. Thus, the nonlinear companding function at the transmitter can be obtained as
\begin{equation}
h_{2}(x)= \begin{cases}
x & \left | x \right |\leq c \sigma\\ 
sgn(x)\frac{1}{k_{1}}\left ( k_{1}c\sigma -\frac{2c}{\sigma } e^{-c^{2}}+\sqrt{\frac{4c^{2}}{\sigma ^{2}}e^{-2c^{2}}+2k_{1}\left (  e^{-c^{2}}-e^{-\frac{x^{2}}{\sigma ^{2}}}\right )}\right ) & \left | x \right |>  c \sigma 
\end{cases}
\end{equation}
At the receiver is the reciprocal of companding transform is
\begin{equation}
h_{2}^{-1}(x)= \begin{cases}
x & \left | x \right |\leq c \sigma\\ 
sgn(x)\sqrt{-\sigma ^{2}ln\left ( -\frac{k}{2}x^{2}+c\left ( k_{1}\sigma-\frac{2}{\sigma e^{-c^{2}}}  \right )x-\frac{k_{1}c^{2}\sigma ^{2}}{2} +e^{-2c^{4}-c^{2}}\right )}& \left | x \right |>  c \sigma 
\end{cases}
\end{equation}
\subsection{Theoretical Analysis}
The Bussgang theorem extension can be used to express real or complex Gaussian signals as the sum of an appropriate compressed input replica and an uncorrelated nonlinear signal distortion noise \cite{dardari2000theoretical}. Hence, the reconstructed signal $y_{n}$ can be formed as
\begin{equation}
y_{n}=\alpha x_{n}+u_{n}
\end{equation}
here the FBMC signal $x_{n}$ is not stationary because of $\alpha$ is attenuation factor \cite{banelli2003theoretical} and the FBMC signal promises $\alpha$  to be time invariant given as $\alpha=\frac{E\left \{ y_{n}x_{n}^{*} \right \}}{E\left \{ x_{n}x_{n}^{*} \right \}}$ and assume a constant average power level throughout the companding process expressed as
%\begin{equation*}
%P_{x_{n}}=P_{y_{n}}=P_{\alpha x_{n}}+P_{u_{n}}=\alpha ^{2}P_{x_{n}}+P_{u_{n}} \Rightarrow P_{u_{n}}=\left ( 1-\alpha ^{2} \right )P_{x_{n}}
%\end{equation*}
\begin{equation}
\begin{split}
P_{y_{n}}&=P_{\alpha x_{n}}+P_{u_{n}}\\
&=\alpha ^{2}P_{x_{n}}+P_{u_{n}} \Rightarrow P_{u_{n}}\\
&=\left ( 1-\alpha ^{2} \right )P_{x_{n}}
\end{split}
\end{equation}

Where, $0 < \alpha < 1$. For $\alpha \rightarrow 1$, the smaller $P_{u_{n}}$ will be the reconstructed signal $x_{n}^{'}$ at the receiver. Which is given by $x_{n}^{'}=\beta y_{n}+z_{n}$, where $\beta =\frac{1}{\alpha }$, $z_{n}=\frac{u_{n}}{\alpha }$. Due to homogeneity, the received signal expressed as $r_{n}=y_{n}+w_{n}$, where $w_{n}$ is the Gaussian channel noise.
A transform gain G is defined as the ratio of the original signal's PAPR to the combined signal's PAPR. 
\begin{equation}
G_{y}dB=10.log_{10}\tfrac{PAPR_x}{PAPR_y}
\end{equation}

\section{SIMULATIONS ANALYSIS}
In this section, simulations are used to evaluate the proposed techniques. Consider 10,000 symbols with 512 sub-carriers for the suggested linear scheme signal transmission.  
The PHYDYAS prototype filter with an overlapping factor of 4 is then applied to the symbols $x_{k}(m)$ \cite{bellanger2010fbmc} . 
Fig. 5 shows how different methods perform in terms of PAPR reduction. The CCDF at $10^{-2}$ , the PAPR 4dB , which is 1.1dB  and  6.3dB smaller than that of the $\mu$ law companding and original signals respectively. A greater reduction can be achieved by proposed linear scheme, which results in 10 dB enhances then the conventional scheme.
 
\begin{figure}[ht]
\center
\includegraphics[scale=0.42]{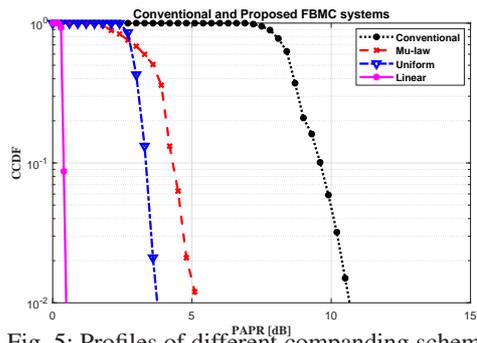}
\vspace{-0.5cm} 
\caption{Profiles of different companding schemes}
\end{figure}
Fig. 5 depicts CCDF curves of various FBMC systems with different length of subcarriers 512 and 1024. When number of subcarriers increasing the PAPR also increasing but in proposed nonlinear scheme is less effected compared with other exiting schemes. Depicts in Table I, to gain a PAPR of 1.0dB the maximum number subcarriers required is 1024. The required PAPRs under the conventional, $\mu$-law, uniform and nonlinear companding schemes are 10.6dB, 5.4dB, 4.9dB and 1.2dB respectively.

\begin{figure}[H]
\center
\includegraphics[scale=0.6]{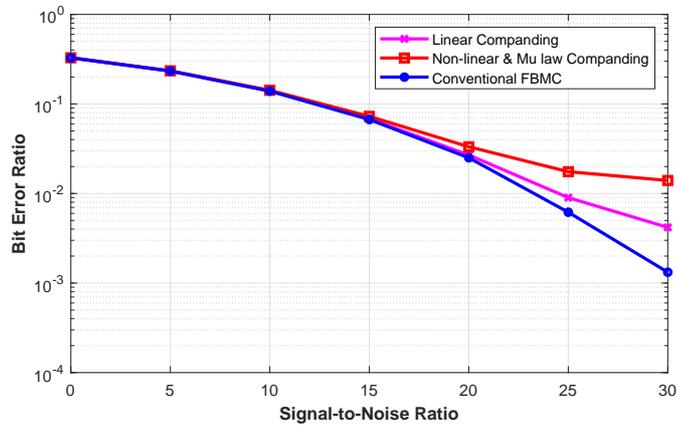}
\caption{BER of different companding schemes}
\end{figure}
Fig. 6 illustrates the BER performance of the conventional and different designed nonlinear FBMC schemes, the AWGN channel is assumed in the simulations. Specially, in Table I, to attain a BER of $10^{-3}$, the minimum signal to noise ratio $(E_{g}/N_{0})$ required is 30dB. The required $E_{g}/N_{0}$ under the conventional, $\mu$-law, uniform and linear companding schemes are 30dB, 40dB, 40dB and 33dB, respectively.
\begin{figure}[H]
\center
\includegraphics[scale=0.6]{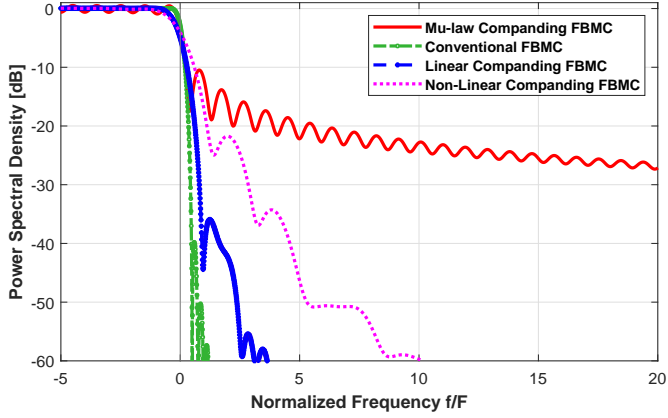}
\caption{PSD of different companding schemes}
\end{figure}

Fig. 7 illustrates the PSD simulation performance of the conventional and different designed nonlinear FBMC schemes. Specially, in Table I, the required PSD under the conventional, $\mu$-law, uniform and linear companding schemes are -40dB, -10dB, -22dB and -37dB, respectively.

\begin{table}[ht]
\centering % used for centering table
    \begin{tabular}{|c|c|c|c|c|c|c|}
    \hline 
    FBMC & \multicolumn{2}{|c|}{SNR (dB)}& \multicolumn{2}{|c|}{PAPR (dB)}& PSD(dB) \\  \cline{2-5}
     Systems&\multicolumn{1}{|c|}{BER=$10^{-2}$}&{BER=$10^{-3}$}& \multicolumn{1}{|c|}{K=512}&K=1024&\\ \hline
    Conventional& 23& 30& 10.3& 10.6 &-40 \\ \hline
    $\mu$-law & 35 & 40 & 5.1 & 5.4&-10 \\ \hline
    Uniform & 35& 40& 4.0& 4.9&-22\\ \hline
    Linear & 25& 33& 1.0& 1.2&-37\\ \hline
    \end{tabular}
    \caption{PAPR, BER and PSD with different companding}
 \end{table}

\section{CONCLUSIONS}
In this paper, the theoretical PAPR expression of an efficient companding FBMC schemes are derived. Through a specially designed nonlinear companding process, the innovative technology maintains a steady average power level. Hence, the efficiency of the HPA can be improved. The proposed nonlinear scheme may meet various scheme criteria by reducing the effective power ratio and improving the BER performance. The proposed schemes are numerous particular examples of the basic $\mu$-law companding, piecewise, and Tangent companding techniques. Additionally, it has been proven through a theoretical analysis that the developed scheme can reduce distortion better than existing schemes. It is further proven that when the slope $k_1$ is modest enough, the de-companding operation at the receiver side is no longer required. The designed linear companding technique might outperform the $\mu$-law companding scheme in terms of power ratio reduction, PSD and BER performance, according to simulation findings. 

\bibliographystyle{unsrt}
\bibliographystyle{plain}
\bibliography{library}

\begin{thebibliography}{10}

\bibitem{samal20195g}
Umesh~Chandra Samal, Bhargav Appasani, and Dusmanta~Kumar Mohanta.
\newblock 5g communication networks and modulation schemes for next-generation
  smart grids.
\newblock In {\em Smart Grids and Their Communication Systems}, pages 361--399.
  Springer, 2019.

\bibitem{jiang2017oqam}
Tao Jiang, Da~Chen, Chunxing Ni, and Daiming Qu.
\newblock {\em OQAM/FBMC for future wireless communications: Principles,
  technologies and applications}.
\newblock Academic Press, 2017.

\bibitem{nissel2017filter}
Ronald Nissel, Stefan Schwarz, and Markus Rupp.
\newblock Filter bank multicarrier modulation schemes for future mobile
  communications.
\newblock {\em IEEE Journal on Selected Areas in Communications},
  35(8):1768--1782, 2017.

\bibitem{bellanger2010fbmc}
Maurice Bellanger, D~Le~Ruyet, D~Roviras, M~Terr{\'e}, J~Nossek, L~Baltar,
  Q~Bai, D~Waldhauser, M~Renfors, T~Ihalainen, et~al.
\newblock Fbmc physical layer: a primer.
\newblock {\em PHYDYAS, January}, 25(4):7--10, 2010.

\bibitem{prasad2004ofdm}
Ramjee Prasad.
\newblock {\em OFDM for wireless communications systems}.
\newblock Artech House, 2004.

\bibitem{cho2010mimo}
Yong~Soo Cho, Jaekwon Kim, Won~Y Yang, and Chung~G Kang.
\newblock {\em MIMO-OFDM wireless communications with MATLAB}.
\newblock John Wiley \& Sons, 2010.

\bibitem{ramavath2012analytical}
Srinivas Ramavath and Rakhesh~Singh Kshetrimayum.
\newblock Analytical calculations of ccdf for some common papr reduction
  techniques in ofdm systems.
\newblock In {\em 2012 International Conference on Communications, Devices and
  Intelligent Systems (CODIS)}, pages 393--396. IEEE, 2012.

\bibitem{an2018design}
Changyoung An and Heung-Gyoon Ryu.
\newblock Design and performance comparison of w-ofdm under the nonlinear hpa
  environment.
\newblock {\em Wireless Personal Communications}, 98(1):983--999, 2018.

\bibitem{he2017filter}
Zongmiao He, Lingyu Zhou, Yiou Chen, and Xiang Ling.
\newblock Filter optimization of out-of-band emission and ber analysis for
  fbmc-oqam system in 5g.
\newblock In {\em 2017 IEEE 9th International Conference on Communication
  Software and Networks (ICCSN)}, pages 56--60. IEEE, 2017.

\bibitem{na2017low}
Dongjun Na and Kwonhue Choi.
\newblock Low papr fbmc.
\newblock {\em IEEE Transactions on Wireless Communications}, 17(1):182--193,
  2017.

\bibitem{zhang2017filtered}
Lei Zhang, Ayesha Ijaz, Pei Xiao, Mehdi~M Molu, and Rahim Tafazolli.
\newblock Filtered ofdm systems, algorithms, and performance analysis for 5g
  and beyond.
\newblock {\em IEEE Transactions on Communications}, 66(3):1205--1218, 2017.

\bibitem{adebisi2018enhanced}
Bamidele Adebisi, Kelvin Anoh, and Khaled~M Rabie.
\newblock Enhanced nonlinear companding scheme for reducing papr of ofdm
  systems.
\newblock {\em IEEE Systems journal}, 13(1):65--75, 2018.

\bibitem{jiang2007nonlinear}
Tao Jiang, Weidong Xiang, Paul~C Richardson, Daiming Qu, and Guangxi Zhu.
\newblock On the nonlinear companding transform for reduction in papr of mcm
  signals.
\newblock {\em IEEE Transactions on Wireless Communications}, 6(6):2017--2021,
  2007.

\bibitem{zakaria2016theoretical}
Rostom Zakaria and Didier Le~Ruyet.
\newblock Theoretical analysis of the power spectral density for fft-fbmc
  signals.
\newblock {\em IEEE Communications Letters}, 20(9):1748--1751, 2016.

\bibitem{bulusu2014reduction}
SS~Krishna~Chaitanya Bulusu, Hmaied Shaiek, Daniel Roviras, and Rafik Zayani.
\newblock Reduction of papr for fbmc-oqam systems using dispersive slm
  technique.
\newblock In {\em 2014 11th International Symposium on Wireless Communications
  Systems (ISWCS)}, pages 568--572. IEEE, 2014.

\bibitem{ji2014semi}
Jinwei Ji, Guangliang Ren, and Huining Zhang.
\newblock A semi-blind slm scheme for papr reduction in ofdm systems with
  low-complexity transceiver.
\newblock {\em IEEE Transactions on Vehicular Technology}, 64(6):2698--2703,
  2014.

\bibitem{li2014partition}
Li~Li, Daiming Qu, and Tao Jiang.
\newblock Partition optimization in ldpc-coded ofdm systems with pts papr
  reduction.
\newblock {\em IEEE Transactions on Vehicular Technology}, 63(8):4108--4113,
  2014.

\bibitem{lu2012sliding}
Shixian Lu, Daiming Qu, and Yejun He.
\newblock Sliding window tone reservation technique for the peak-to-average
  power ratio reduction of fbmc-oqam signals.
\newblock {\em IEEE Wireless Communications Letters}, 1(4):268--271, 2012.

\bibitem{lee2016low}
Kang-Seok Lee, Young-Jeon Cho, Jun-Young Woo, Jong-Seon No, and Dong-Joon Shin.
\newblock Low-complexity pts schemes using ofdm signal rotation and
  pre-exclusion of phase rotating vectors.
\newblock {\em Iet Communications}, 10(5):540--547, 2016.

\bibitem{deng2017modified}
Honggui Deng, Shuang Ren, Yan Liu, and Chengying Tang.
\newblock Modified pts-based papr reduction for fbmc-oqam systems.
\newblock In {\em Journal of Physics: Conference Series}, volume 910, page
  012057. IOP Publishing, 2017.

\bibitem{ye2013papr}
Chen Ye, Zijun Li, Tao Jiang, Chunxing Ni, and Qi~Qi.
\newblock Papr reduction of oqam-ofdm signals using segmental pts scheme with
  low complexity.
\newblock {\em IEEE Transactions on Broadcasting}, 60(1):141--147, 2013.

\bibitem{van2014papr}
Nuan Van~der Neut, Bodhaswar~TJ Maharaj, Frederick De~Lange, Gustavo~J
  Gonz{\'a}lez, Fernando Gregorio, and Juan Cousseau.
\newblock Papr reduction in fbmc using an ace-based linear programming
  optimization.
\newblock {\em EURASIP Journal on Advances in Signal Processing},
  2014(1):1--21, 2014.

\bibitem{yang2005ace}
Zhixing Yang, Haidong Fang, and Changyong Pan.
\newblock Ace with frame interleaving scheme to reduce peak-to-average power
  ratio in ofdm systems.
\newblock {\em IEEE Transactions on Broadcasting}, 51(4):571--575, 2005.

\bibitem{liu2019low}
Zilong Liu, Pei Xiao, and Su~Hu.
\newblock Low-papr preamble design for fbmc systems.
\newblock {\em IEEE Transactions on Vehicular Technology}, 68(8):7869--7876,
  2019.

\bibitem{cui2019peak}
Fangyu Cui, Yunlong Cai, Minjian Zhao, Ming Lei, and Lajos Hanzo.
\newblock Peak-to-average power ratio reduction based on penalty-cccp for
  filter bank multicarrier systems.
\newblock {\em IEEE Transactions on Vehicular Technology}, 68(11):11353--11357,
  2019.

\bibitem{ramavath2021theoretical}
Srinivas Ramavath and Umesh~Chandra Samal.
\newblock Theoretical analysis of papr companding techniques for fbmc systems.
\newblock {\em Wireless Personal Communications}, 118(4):2965--2981, 2021.

\bibitem{chunkath2021constrained}
Job Chunkath and VS~Sheeba.
\newblock Constrained message length coding for low peak to average power ratio
  in fbmc: Oqam systems.
\newblock {\em Wireless Personal Communications}, 116(4):2981--2996, 2021.

\bibitem{geetha2020performance}
MN~Geetha and UB~Mahadevaswamy.
\newblock Performance evaluation and analysis of peak to average power
  reduction in ofdm signal.
\newblock {\em Wireless Personal Communications}, 112(4):2071--2089, 2020.

\bibitem{srivastava2020hybrid}
Mohit~Kumar Srivastava, Manoj~Kumar Shukla, Neelam Srivastava, and Ashok~Kumar
  Shankhwar.
\newblock A hybrid scheme for low papr in filter bank multi carrier modulation.
\newblock {\em Wireless Personal Communications}, 113(2):1009--1028, 2020.

\bibitem{kondamuri2020non}
Shri~Ramtej Kondamuri and Anuradha Sundru.
\newblock Non linear companding transform to mitigate papr in dct based sc-fdma
  system.
\newblock {\em Wireless Personal Communications}, pages 1--20, 2020.

\bibitem{dardari2000theoretical}
Davide Dardari, Velio Tralli, and Alessandro Vaccari.
\newblock A theoretical characterization of nonlinear distortion effects in
  ofdm systems.
\newblock {\em IEEE transactions on Communications}, 48(10):1755--1764, 2000.

\bibitem{banelli2003theoretical}
Paolo Banelli.
\newblock Theoretical analysis and performance of ofdm signals in nonlinear
  fading channels.
\newblock {\em IEEE Transactions on Wireless Communications}, 2(2):284--293,
  2003.

\end{thebibliography}

\end{document}